**Dynamical Spin Injection into p-type Germanium at Room Temperature**

Mariko Koike,[1] Eiji Shikoh,[1] Yuichiro Ando,[1] Teruya Shinjo,[1] Shinya Yamada,[2] Kohei Hamaya,[2] Masashi Shiraishi[1,*]

[1]Department of Systems Innovation, Graduate School of Engineering Science, Osaka University, Toyonaka 560-8531, Japan

[2]Department of Electronics, Kyushu University, 744 Motooka, Fukuoka 819-0395, Japan

* Corresponding author: shiraishi@ee.es.osaka-u.ac.jp

**Abstract**

We demonstrate dynamical spin injection into p-type germanium (Ge) at room temperature (RT) using spin pumping. The generated pure spin current is converted to a charge current by the inverse spin-Hall effect (ISHE) arising in the p-type Ge sample. A clear electromotive force due to the ISHE is detected at RT. The spin-Hall angle, $\theta_{\mathrm{SHE}}$, for p-type Ge is estimated to be ca. $1 \times 10^{-3}$ at RT.

Group-IV spintronics using carbon, silicon (Si) and germanium (Ge), has attracted considerable attention in recent years.[1,2] Carbon and Si essentially exhibit a small spin–orbit interaction (SOI), which allows long coherence of spins injected into these materials. Ge has a comparatively large SOI and exhibits high carrier mobility, enabling coherent spin transport in field effect transistors (FETs) with a short channel. Currently, Ge-based metal-oxide-semiconductor (MOS) FETs are a potential candidate for overcoming the scaling limit of Si-based MOS FETs.[3] A great deal of effort has been undertaken to realize Ge based MOS transistors, including attempts at fabricating high quality Ge on insulators and investigations of metal/Ge contacts.[4-6] In addition, Ge-based spintronics have been studied intensively.[7,8] Although demonstrations of spin injection in Ge at RT have been reported by several groups,[9,10] the measurement techniques were limited mainly to the electrical non-local 3-terminal (NL3T) method. As reported previously, this method does not always allow precise investigations,[11] and a new method for confirming successful spin injection into Ge is eagerly awaited. A promising example is the study by Jain et al.,[12] which used a dynamical method; however, only spin injection into n-type Ge was realized. Since the carrier mobility in p-type Ge is much higher than that in the other p-type semiconductors such as p-Si and p-GaAs, which is significantly superior to any other semiconductors, p-type Ge is expected to be the most promising materials for realization of p-MOS FET. Therefore successful and reliable spin injection in p-type Ge at RT is also strongly desired for establishing Ge-based spintronics. In this letter, we demonstrate spin injection into p-type Ge at RT using a dynamical spin pumping method. The appearance of the inverse spin Hall effect (ISHE) in p-type Ge provides strong evidence of spin injection, and the spin Hall angle for the p-type Ge is estimated.

Figure 1(a) shows a schematic illustration of a sample used in this study. A p-type Ge layer, of which dopant and doping concentration was boron (B) and $1\times10^{18}$ cm$^{-3}$, was formed by using of ion implantation on a Ge substrate of 350 μm in thick, and the thickness of the doped layer was set to be 100 nm. A 25-nm-thick Ni$_{80}$Fe$_{20}$ (Py) layer was formed on the p-type Ge layer by electron beam evaporation at RT. A thin Al layer (~4 nm) was formed without breaking the vacuum after the Py deposition in order to prevent oxidation of the Py surface. It is worth noting that the Py/p-type Ge contact is ohmic because of Fermi level pinning[13, 14] and that dynamical spin injection is not impeded by the Schottky barrier. In order to inject spins dynamically, we employed a spin pumping method using an electron spin resonance (ESR) system. The details of the spin pumping are described in the literature.[15] During the measurements of ferromagnetic resonance (FMR) signals and output voltages, the samples were placed at the center of a TE$_{011}$ microwave cavity with a frequency of $f$ = 9.1 GHz. In order to excite FMR in the Py layer, an external magnetic field, $H$, was applied to the sample at an angle of $\theta_H$ as shown in Fig. 1(a). In the FMR condition, the spin current density, $j_S$, generated by the spin pumping at the Py/p-type Ge interface is theoretically expressed as[16]

$$j_S = \frac{g_r^{\uparrow\downarrow}\gamma^2 h^2 \hbar \left[4\pi M_S \gamma + \sqrt{(4\pi M_S)^2 \gamma^2 + 4\omega^2}\right]}{8\pi\alpha^2\left[(4\pi M_S)^2 \gamma^2 + 4\omega^2\right]}, \quad (1)$$

where $h$, $\hbar$, and $\alpha$ are the microwave magnetic field, the Dirac constant and the Gilbert damping constant, respectively. $\omega$ (=$2\pi f$, where $f$ is the microwave frequency) is the angular frequency of the magnetization precession. $g_r^{\uparrow\downarrow}$ is the real part of the mixing conductance and is given by[16]

$$g_r^{\uparrow\downarrow} = \frac{2\sqrt{3}\pi M_S \gamma d_F}{g\mu_B \omega}\left(W_{F/N} - W_F\right), \tag{2}$$

where $g$, $\mu_B$, $d_F$, $W_{F/N}$, and $W_F$ are the $g$ factor, the Bohr magneton, the thickness of the Py layer, the FMR spectral width for the Py/p-type Ge film, and the FMR spectral width for the Py film, respectively. The generated spin current diffuses from the Py/Ge interface into the Ge layer. The injected spins are converted to a charge current due to the ISHE, resulting in the generation of an electric voltage. The electromotive force due to the ISHE is expressed as[16]

$$V_{ISHE} = \frac{w\theta_{SHE}\lambda_N \tanh(d_N/2\lambda_N)}{d_N\sigma_N + d_F\sigma_F}\left(\frac{2e}{\hbar}\right)j_S, \tag{3}$$

where $w$ is the length of the Py layer defined as in Fig. 1(a). $d_F$ and $\sigma_F$ are the thickness and electric conductivity of the Py layer, respectively. $d_N$ and $\sigma_N$ are the thickness and electric conductivity of the p-type Ge layer, respectively. Since the Ge substrate is not insulative enough (see Fig. 1(a)), the contribution from the B-doped channel and the substrate to the total resistance of the sample should be taken into account for precise analyses of the results, and the measured conductivity was used for the calculation of the spin Hall angle.

Figure 1(b) shows the FMR spectra, $dI(H)/dH$, for the Py/p-type Ge/Ge(111) sample (red curve) and the Py/SiO$_2$/Si(100) sample (black curve) under a microwave excitation power of $P_{MW}$ = 200 mW for $\theta_H$ = 0°. Here, $I$ denotes the microwave absorption intensity. For the Py/p-type Ge sample, $W_{F/N}$ was estimated to be 2.97 mT, which is clearly larger than that for the Py/SiO$_2$/Si(100) sample ($W_F$ = 2.56 mT). It was confirmed that the enhancement of the spectral width was reproduced, although the enhancement was not strong. Since the spin angular momentum of the Py layer is reduced due to the generation of the spin current in the Ge channel in the spin pumping condition, the increase in $W$ can be regarded as evidence of spin pumping, as expressed in Eqs. (1) and (2).

Further evidence for spin injection is provided in the following discussion. Figure 1(c) shows the dc electromotive force signals (open circles), $V_{out}$, as a function of $H$ for the Py/p-type Ge sample under a microwave excitation power of $P_{MW}$ = 200 mW. A clear output voltage, $V_{out}$, of about 10 μV was observed under FMR. In order to eliminate the contribution from the anomalous Hall effect (AHE), the following equation is used:[17]

$$V = V_{ISHE} \frac{\Gamma^2}{(H-H_{FMR})^2 + \Gamma^2} + V_{AHE} \frac{-2\Gamma(H-H_{FMR})}{(H-H_{FMR})^2 + \Gamma^2}, \quad (4)$$

where $\Gamma$ denotes the damping constant and $H_{FMR}$ is the ferromagnetic resonance field (93.8 mT for $\theta_H$ = 0° in this study). The first term, which has a symmetrical Lorentzian shape, corresponds to the contributions from the ISHE in the p-type Ge, whereas the second term, which has an asymmetrical shape, contains the AHE in the Py film.[16,17] As shown in Fig. 1(c), a theoretical fit using Eq. (4) nicely reproduces the experimental results, which also supports the conclusion that successful spin injection occurs. $V_{out}$–$H$ curves under various microwave powers and the magnitude of the $V_{ISHE}$ as a function of $P_{MW}$ are shown in Figs. 2(a) and 2(b), respectively. As expressed in Eqs. (1)–(3), $V_{ISHE}$ has an $h^2$ dependence, where $h$ is the microwave magnetic field. Since $h$ has a linear relationship with $\sqrt{P_{MW}}$, $V_{ISHE}$ is expected to increase linearly with increasing $P_{MW}$. As can be seen in Fig. 2(b), the $P_{MW}$ dependence of $V_{ISHE}$ is in good accordance with the theory, indicating that the observed electromotive forces can be attributed to the ISHE of the p-type Ge, due to the fact that spin injection successfully occurs. The other important conclusion from the experiment is that the FMR spectra does not saturate, which enables an estimation of the spin Hall angle for p-type Ge. Furthermore, the angular dependence of the electromotive forces from the p-type Ge also corroborates our claim that the inverse spin Hall effect occurs. The FMR spectra and the $V_{out}$–$H$ curves for the Py/p-type Ge sample for various $\theta_H$ are displayed in Figs. 3(a) and 3(b), respectively. With changing $\theta_H$, the polarity of the $V_{out}$–$H$ curves changes across $\theta_H = 90°$, which is consistent with the symmetry of the ISHE, $E_{ISHE} \sim J_S \times \sigma$. For $\theta_H = 90°$, no output voltage can be seen despite the clear FMR spectrum. This result is also consistent with the theory for the ISHE.

From the results and discussion above, we can conclude that spin injection into p-type Ge using spin pumping and conversion of a pure spin current to a charge current due to the ISHE are successfully demonstrated at RT. We now estimate the spin Hall angle, $\theta_H$, for p-type Ge. Using Eqs. (1)–(3), with the parameters shown in Table I, the real part of the mixing conductance and the spin current density at the Py/p-type Ge interface are calculated to be $g_r^{\uparrow\downarrow} = 1.8 \times 10^{19}$ m$^{-2}$ and $j_S = 8.4 \times 10^{-9}$ Jm$^{-2}$, respectively. $h$ was calculated to be 0.16 mT by using a conventional method [18], $\lambda_N$ was estimated to be ca. 30 nm from the spin lifetime for holes in bulk Ge at RT (700 fs) [19] and the diffusion constant in Ge with the hole concentration of ca. $1 \times 10^{18}$ cm$^{-3}$ (ca. 10 cm$^2$/s). [20] Using these values and $V_{ISHE} = 13.5$ μV, the spin-Hall angle, $\theta_{SHE}$, for p-type Ge film is estimated to be ca. $1 \times 10^{-3}$, which is larger than that of p-type Si.[21]

In summary, we demonstrated successful spin injection into p-type Ge using dynamical spin pumping at RT and experimentally corroborated this demonstration by observing the electromotive forces due to the ISHE of p-type Ge. The spin Hall angle, $\theta_{SHE}$, for p-type Ge was estimated to be ca. $1 \times 10^{-3}$, which is larger than that of p-type Si [21].

This study is partly supported by Toray Sci. Foundation and Grant-in-Aid for Scientific Research (KAKENHI).

Figure captions

Figure 1 | **Sample structure and electromotive forces in p-type Ge.**

(a) Schematic illustration of the Py/p-type Ge sample. $H$ represents the external magnetic field and $\theta_H$ is the magnetic field angle to the $Ni_{80}Fe_{20}$ film plane. The dimensions of the sample are described in the figure.

(b) Magnetic field ($H$) dependence of the FMR signals, d$I(H)$/d$H$, for the Py sample (black line) and Py/p-type Ge sample (red line), for $\theta_H = 0°$ at RT. The ferromagnetic resonance fields, $H_{FMR}$, for the Py sample and Py/p-type Ge sample are 102 and 93.8 mT, respectively. The expanded graphs around the peaks of d$I(H)$/d$H$ are shown in the insets.

(c) Magnetic field dependence of the electromotive force, $V_{out}$, for the Py/p-type Ge film at RT. The open circles and red solid line show the experimental data and fitting result, respectively. Deconvolution of the ISHE signal and AHE signal is successfully achieved. The blue solid line and dashed line show only the first term contribution of $V_{ISHE}$ and only the second term contribution of $V_{AHE}$ in Eq. (4), respectively.

Figure 2 | **Microwave power dependence of the electromotive forces in p-type Ge.**

(a) Magnetic field ($H$) dependence of the electromotive forces for the Py/p-type Ge sample under various microwave excitation powers at $\theta_H = 0°$.

(b) Microwave power dependence of the $V_{ISHE}$ for the Py/p-type Ge sample, where $V_{ISHE}$ was estimated from a fit using Eq. (4). The solid lines show a linear fit to the data.

Figure 3 | **Angular dependence of ISHE signal.**

Magnetic-field-angle, $\theta_H$, dependence of **(a)** the FMR spectra, d$I(H)$/d$H$, and **(b)** the electromotive forces, $V_{out}$ for the Py/p-type Ge sample. Apparent reversal of the electromotive forces can be seen.

**Table I | Physical parameters for estimating the spin Hall angle.**

Magnetic and electrical parameters for the Py/p-type Ge sample used for the theoretical estimation of the spin Hall angle. The g-factor and γ are referred from Refs. 22 and 18, respectively.

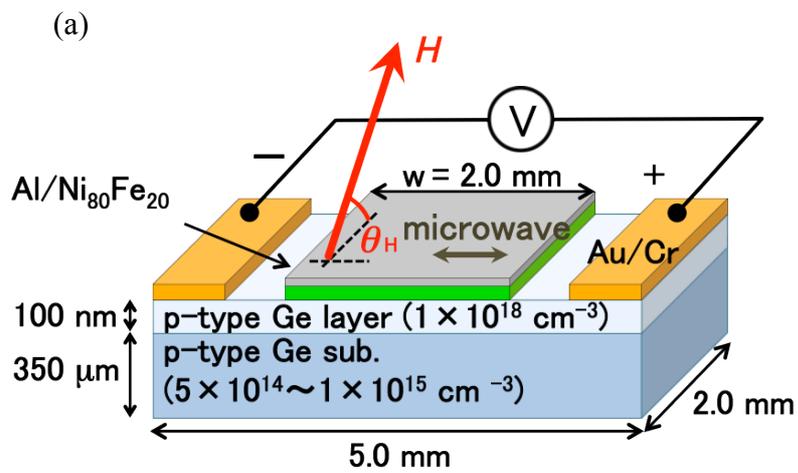

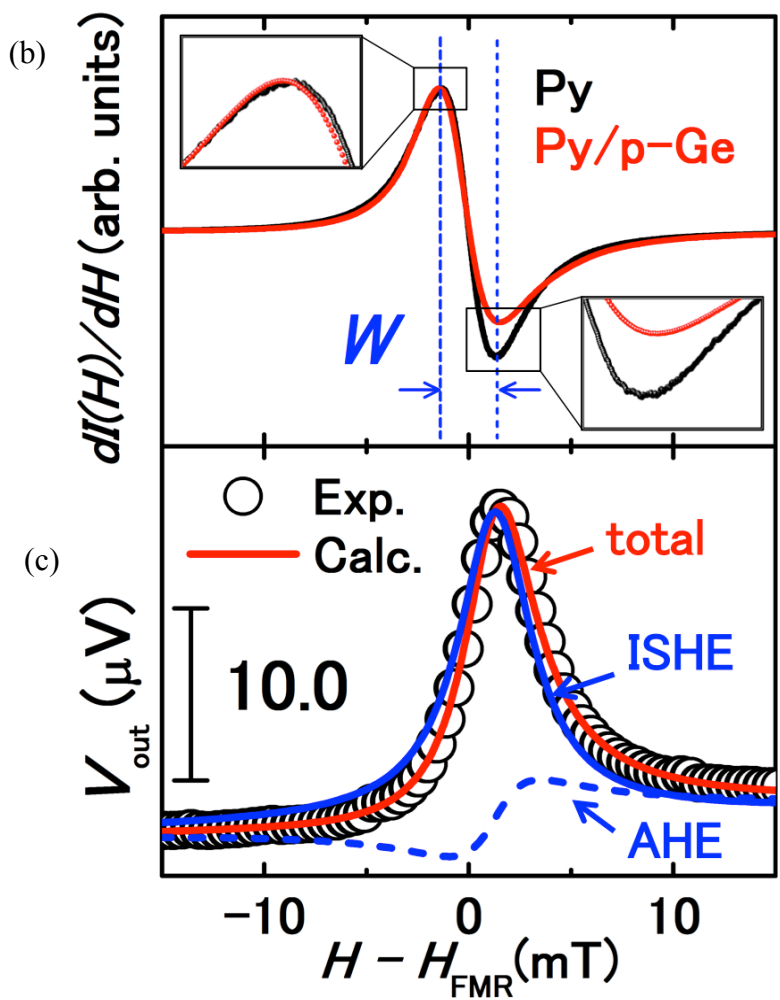

M. Koike *et al.,* Figure 1 Sample structures and results of ISHE in p-type Ge.

(a) 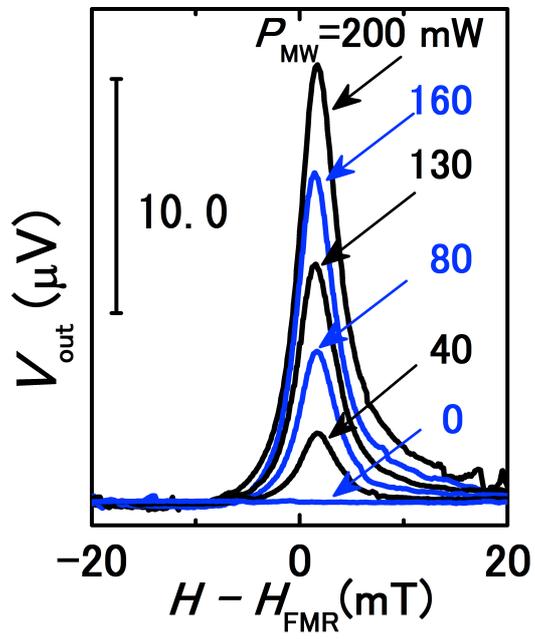 (b) 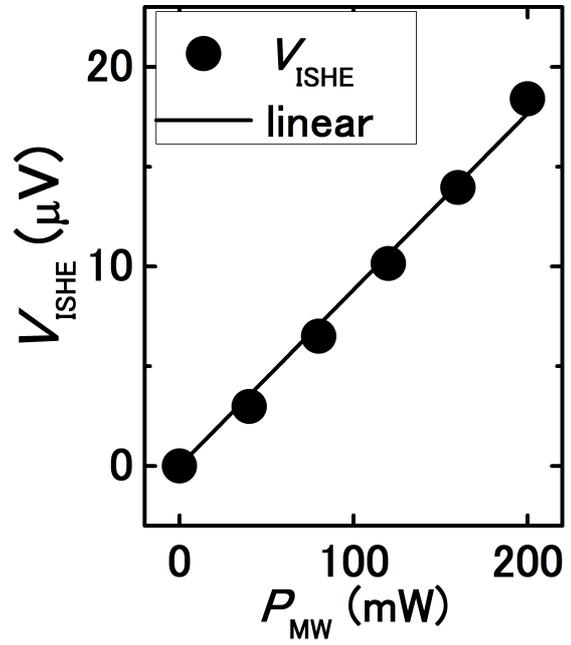

M. Koike *et al.*, Figure 2 Microwave power dependence of the electromotive force in the p-type Ge.

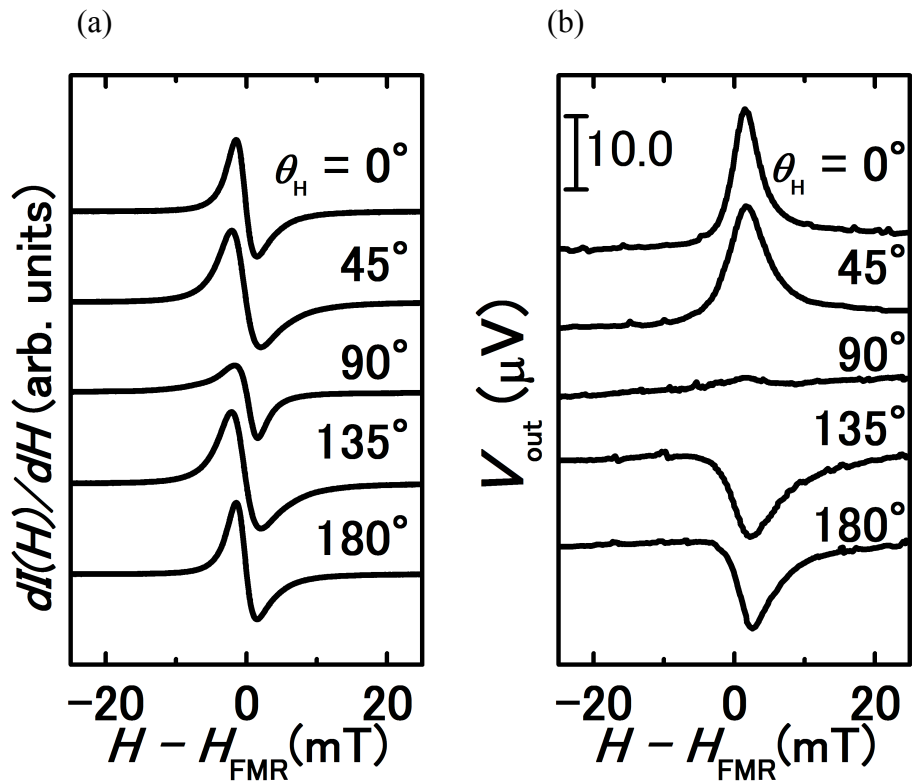

M. Koike *et al.,* Figure 3 Angular dependence of the ISHE signals.

| g-Factor Ref. 21 | $4\pi M_S$ [T] | $\gamma$ [T$^{-1}$s$^{-1}$] Ref. 18 | $\omega$ [s$^{-1}$] | $\alpha$ | $W_F$ [mT] | $W_{F/N}$ [mT] | $d_F\sigma_F$ [$\Omega^{-1}$] | $d_N\sigma_N$ [$\Omega^{-1}$] | w [mm] |
|---|---|---|---|---|---|---|---|---|---|
| 1.6 | 0.918 | 1.86×10$^{11}$ | 5.73×10$^{10}$ | 0.008 | 2.56 | 2.97 | 0.063 | 0.036 | 2 |

M. Koike *et al.,* Table1

The parameters.